# NEED-DRIVEN DECISION-MAKING AND PROTOTYPING FOR DLT: FRAMEWORK AND WEB-BASED TOOL

*Research Paper*

Tomas Bueno Momčilović, fortiss, Germany, momcilovic@fortiss.org
Matthias Buchinger, fortiss, Germany
Dian Balta, fortiss, Germany, balta@fortiss.org

## Abstract

*In its 14 years, distributed ledger technology has attracted increasing attention, investments, enthusiasm, and user base. However, ongoing doubts about its usefulness and recent losses of trust in prominent cryptocurrencies have fueled deeply skeptical assessments. Multiple groups attempted to disentangle the technology from the associated hype and controversy by building workflows for rapid prototyping and informed decision-making, but their mostly isolated work leaves users only with fewer unclarities. To bridge the gaps between these contributions, we develop a holistic analytical framework and open-source web tool for making evidence-based decisions. Consisting of three stages – evaluation, elicitation, and design – the framework relies on input from the users' domain knowledge, maps their choices, and provides an output of needed technology bundles. We apply it to an example clinical use case to clarify the directions of our contribution charts for prototyping, hopefully driving the conversation towards ways to enhance further tools and approaches.*

## 1      Introduction

Distributed ledger technology (DLT), including blockchain, is a set of promising solutions for decentralizing and automating trustworthy transactions. Initially, Bitcoin (Nakamoto, 2008) and Ethereum (Buterin, 2013) propelled the image of blockchain as a general-purpose technology, promising to couple "trustless" relationships with efficiency and investments. Yet, even after a decade of development, many of these promises are still missing value propositions, preventing an otherwise enthusiastic user base from adopting the technology (Deloitte, 2021). In addition to the recent financial tremors that led investors to lose faith in some of the more prominent cryptocurrencies (see, e.g., ), few projects have proven the value of using DLT over traditional databases, digital signatures or trusted third parties, bringing about the sentiment that it is just an expensive experiment (cf. Suichies, 2015; Weaver, 2022).

To move the discussion beyond the controversy and towards a clearer assessment, three groups of methods for supporting decisions emerged. The first group, consisting both of academic (e.g. Treiblmaier, 2019) and industry-based (e.g. Tsoniotis et al., 2019) actors, embraced prototyping as a means of making the technology more tangible. By providing a lower-cost testing environment or a workflow to establish it, this method makes it easier to compare DLT with existing solutions. The second broader group focused on decision trees (e.g., Koens and Poll, 2018) and surveys (e.g., Gourisetti et al., 2019). Using their tools, questions, and additional input, users are driven to make informed decisions before developing a DLT-based system themselves. The third group, consisting mostly of contributors to technical journals and repositories, focused almost-exclusively on differentiating, improving, and extending distributed ledger architectures, to make them more attractive from a technical standpoint. Such engineering initiatives reduced the weaknesses (e.g., faster consensus mechanisms; Xiao et al., 2020) to highlight the strengths, effectively moving the discussion away from the traditional blockchain architecture (e.g., directed acyclic graphs instead of single chains of blocks; Baird and Luykx, 2020) towards a larger set of distributed solutions with similar or practically same properties.

Although these three groups once again broadened the expectations of what the technology can do (cf. e.g., Litan, 2021), their methods were pursued mostly independently. Our initial approach connected them into one framework (Bueno Momčilović et al., 2022), combining informed decision-making, early prototyping, and recommendations of architectural components based on their properties, not traditional approaches or popular apps. However, connections are not sufficient; despite an increasing number of easy-to-use apps and the evident benefits of using prototypes to determine value before investing, we recognize that DLT is often still difficult to implement without strong technical knowledge, but also imprudent to pursue without deep knowledge of the problem domain. Our goal has been to have a one-stop-shop workflow and environment that enables users to combine managerial and technical perspectives, record their choices, and make decisions based on the needs or problems of their use case (i.e., need-driven decisions) rather than the needs or problems of the solution itself. We achieve this by building an open-source tool from the framework, and contribute with an empirical and holistic update to the state-of-the-art.

This paper first shortly introduces the background of our work. It then provides an account of our research approach: the empirical evaluation of prototypes that domain experts and technical developers created collaboratively; a meta-analysis of existing decision frameworks; and a survey of technical literature. We discuss similar attempts to do so in recent years and introduce a new (online) tool with an example use case. Finally, we discuss the wider implications of early prototyping, and strive to move the further discussion towards building minimum viable applications and managing change.

## 2 Background

DLT is a relatively novel paradigm for storing transactions on a commonly accessible ledger without the possibility of changing their content afterwards. Relying on graph theory, linked lists, and hashing - among a variety of computer scientific and economic concepts - the ledger connects each new transaction to the previous ones, thereby creating a shared truth between parties (Wattenhofer, 2017). Given these properties, DLT solutions are meant to change the ways stakeholders form relationships with each other, by having them rely on the algorithms rather than intermediaries or interpersonal trust to establish consensus (Kannengießer et al. 2019). In the DLT context, consensus refers to the agreement between participants that the present state of transactions on the ledger represents the truth, which has been the dominant 'trustless' approach to mitigating fraudulent transactions and mistakes in the code.

Several waves of DLT variations have emerged since the first Bitcoin white paper (Nakamoto, 2008). The most prominent variation relies on the same blockchain-based mechanism, which chains blocks of transactions using cryptographic hash functions (Burkhardt et al., 2018). These traditional approaches use the costly Proof-of-Work algorithms to build consensus among freely joining participants, mostly by incentivizing users to solve increasingly complex equations by committing computational resources and competing between each other for the associated reward. The second wave of variations relied on the blockchain, but instead proposed using different consensus algorithms such as Proof-of-Stake or Proof-of-Authority (Xiao et al., 2020). These algorithms attach more weight to the participants themselves, whose transaction writing permissions may be more-or-less constrained, rather than any entrant's ability to commit resources to solve enormously and increasingly costly mathematical puzzles (as of November 2022; cf. e.g., Jones et al., 2022; Duggan, 2022).

The most recent but less prominent wave of variations is instead based on extensions, more flexible configurations, and different ways of linking transactions. Regarding extensions, apps have become software platforms with a list of optional on-chain components (i.e., based on DLT) and off-chain extensions (i.e., connected non-DLT apps; Fridgen et al., 2018). For example, smart contracts, or code within the DLT network that automatically executes transactional events once pre-specified contractual terms are fulfilled (Mills et al. 2016), is what especially distinguishes the newer wave from other alternatives (Duivestein et al. 2015). Regarding configurations, an increasing number of apps offer an entire spectrum of access controls, in three general areas: **public** writing and reading permissions; private writing but public reading permissions (a.k.a. **permissioned** DLT); and **private** writing and reading permissions. Regarding linking mechanisms, some may be much older than the blockchain itself (cf. e.g., Weaver, 2022). Innovations using directed acyclic graphs (e.g., hashgraphs; Baird and Luykx, 2020), for example, have recently come to the fore. This expansion of possibilities contributes to the difficulty in deciding which solution to use.

# 3 Research Approach

## 3.1 Sources and Methods

To provide a framework that supports decisions and early prototyping, we conducted an exploratory qualitative analysis of existing theory and practice. We developed a descriptive artefact in the form of an analytical framework, which can be categorized as a theory for analyzing[1] (Gregor, 2006). The analysis is rooted in the paradigm of pragmatism (i.e., constructive knowledge; Goldkuhl, 2012) within the argumentative-deductive set of approaches (i.e., formal and linguistic methods underlying systematic literature reviews; Wilde & Hess, 2007). Our review methodology has been based on the iterative approach to reviewing documents, texts, and other bodies of knowledge such as repositories systematically, also referred to as the hermeneutic review (Boell and Cecez-Kecmanovic, 2014), iterative analysis (Srivastava and Hopwood, 2009), and grounded theory in non-technical circles (Glaser and Strauss, 1967).

We reviewed the following accessible body of knowledge (cf. Figure 1 for steps): steps 0-1 involving 19 case studies and prototypes that focused on solving real-life challenges (cf. Table 1), elaborated by graduate students (2-6 member teams) alongside partners from practice in fortiss (2023); step 2 involving 36 decision frameworks for evaluating DLT use cases; and step 3 involving 32 technical papers focused on architecture design. We address the inputs, the methods, and the outputs of each source according to the stages they contributed to, and discuss step 4 in the next section (cf. Figure 2).

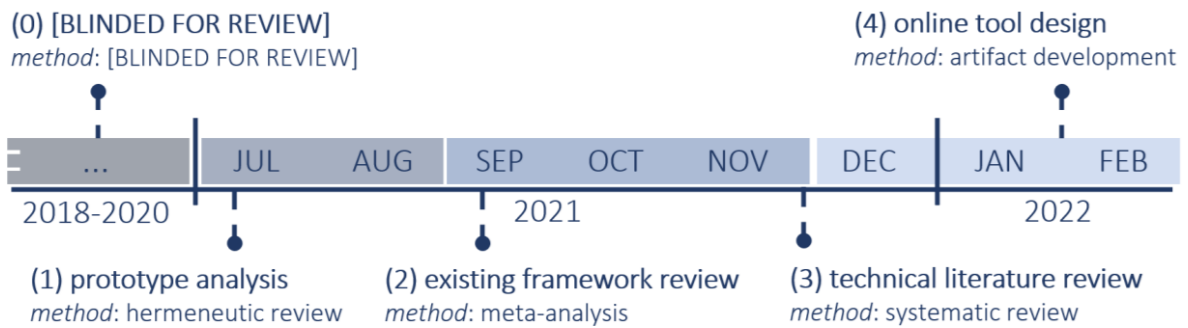

*Figure 1.    Timeline of research and framework development*

To study the decision frameworks comprehensively, we surveyed 289 articles from journals, books, conferences, and white papers. Papers were initially retrieved using search queries via Scopus and other academic search engines, and the list was then expanded using the snowball method of mining references and citations (Wohlin, 2014). Our search queries included a combination of keywords related to DLT (e.g., distributed ledger, blockchain, off-chain, etc.) and decision frameworks (e.g., decision tree, framework, workflow, questionnaire, etc.). We found 107 decision trees, questionnaires, and workflows in the literature, together comprising more than 400 questions.

We relied on two approaches from related work to analyze them systematically (Shepperd, 2013) and exclude subjective, informal, or brainstorming examples where no unique contributions could be found. First, by applying the approach developed by Koens and Poll (2018), we extracted all questions and outcomes from the existing frameworks, and then clustered them according to similarity. Second, we used the list developed by Colomo-Palacios et al. (2020), who studied all unique DLT-related criteria from technical and business literature, to refine our clusters into mutually exclusive groups and add more. Following the approach taken in the frameworks we analyzed, we formulated one question out of each resulting group and connected the questions with the most frequent outcomes (or multiple, where necessary). With this, we designed two stages in the framework: the **evaluation** stage which helps users to determine whether their use case and DLT are compatible; and the **elicitation** stage which prompts users to determine their needs and record them in the choice of the three most access control configurations. The early choice of access control setting is an artifact of many existing frameworks, which we retain here, but with the caveat that questions from other sources with more detailed design-oriented answers may contradict these access control outcomes of existing frameworks.

---

[1] As defined by Gregor (p. 619, 2006): „a description of the phenomena of interest, analysis of relationships among those constructs, the degree of generalizability in constructs and relationships and the boundaries within which relationships, and observations hold."

| *Existing frameworks* | | | | *Prototypes* | | | |
|---|---|---|---|---|---|---|---|
| **Types** | **%** | **Stages** | **%** | **Incentives** | **%** | **DLT access** | **%** |
| Decision trees | 81 | Evaluation | 97 | Compliance | 63 | Private | 53 |
| Questionnaires | 15 | Elicitation | 75 | Financial | 58 | Permissioned | 32 |
| Other | 4 | Design | 8 | Prosocial | 16 | Public | 16 |

*Table 1.    Analyzed frameworks prototypes*

To study the cases, we reviewed the documentation from the fortiss (2023) including presentations, reports, demos, and repositories. The background of the prototyping process is the following. All teams started by implementing the Balta et al. (2015) three-step method of need analysis, feasibility study, and architecture design to gather information from the partners, then iteratively built a fully functional prototype with additional input, and then presented their work with discussions of lessons learned. Examples of cases include private DLTs for increasing compliance (e.g. in parts manufacturing, Habes et al., 2019), and a public DLT for incentivizing prosocial behavior (e.g., with tokens, Degenhart et al., 2019; for all cases, cf. fortiss, 2023).

As the documentation involves different types of input – some textual, other visual or code-based – we qualitatively encoded (Srivastava and Hopwood, 2009) the properties of each case as we went through the files. The general categories of properties included (i) the background of the challenge, the actors, their incentives, and similar; and (ii) the design decisions of students in agreement with the partners in the prototyping phase. The outcome we developed included the properties from two categories that most frequently intersected, such as the needs of the private sector to protect intellectual property and the use of private DLT solutions. Since students were attending a DLT-oriented lab, evaluation was not recorded in the documentation, but there has been much useful input for the elicitation and **design** stages - i.e., the detailed translation of business needs into technical requirements for generating component recommendations.

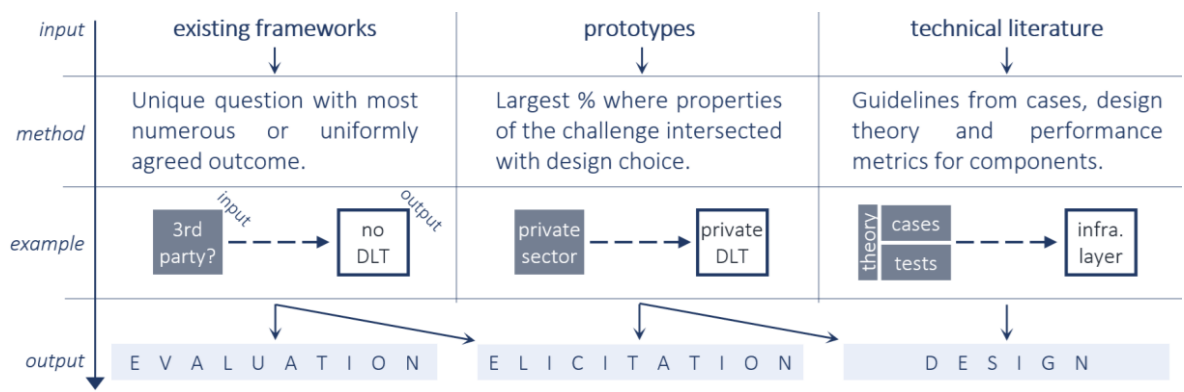

*Fig. 2        Process of generating the questions for the three stages*

To study technical papers, we once more used systematic methods to retrieve them (Wohlin, 2014) and analyze their contents (Shepperd, 2013). This set of literature consists of surveys, comparisons, and recommendations of DLT-based and compatible non-DLT components (cf. e.g., Xu et al., 2017; Kannengießer et al., 2019). We relied on them to formulate nine central aspects in the design stage, deriving component-specific questions from the input within each aspect, and further specifying options and corresponding architectures as output. We suggest suitable approaches or technologies for each component in detail, but also describe their interplay on a level high enough not to depend on current apps.

## 3.2    Related Work

With respect to existing research which similarly combined multiple sources into decision support systems, we highlight three aspects which distinguish our work. First, earlier approaches, including those that build on and combine many existing frameworks, select which questions are important based on their own knowledge and experience (i.e., many existing frameworks mentioned above) or validation by experts (cf. e.g. Betzwieser et al., 2019). Some also build on informal sources (e.g. Twitter), which in our analysis only added one unique but ultimately broad question: do you have a real business case (cf. e.g. Lewis, 2016). By contrast, we found that the formal review served the purpose of quality control, after which we also reviewed and clustered the questions according to their topical similarities, but did not exclude any unique question in case their value for evaluation or elicitation may be proven in a subset of use cases.

Second, to our knowledge and except for very few examples (e.g., Xu et al., 2017), existing frameworks do not provide clear architectural component recommendations, nor bind them to the needs of a use case. Besides open-ended questions of workflows, close-ended examples only offer high-level access control outcomes or suggest specific applications (e.g., Hyperledger Fabric). We instead relied on prototypes and technical literature to determine component suggestions, and strived to keep our contribution independent from the current wave of preferred DLT and non-DLT apps.

Third, three holistic approaches similarly contributed with a combination of evaluation, elicitation and design steps. Betzwieser et al. (2019), for example, provide a general workflow for architecture design with a mixture of open- and close-ended outcomes; Gourisetti et al. (2019) provide close-ended outcomes for picking consensus mechanisms (e.g., Proof-of-Stake); and Abdo & Zeadally (2020) provide ways to iteratively evaluate DLT suitability after making informed decisions, in the face of changing organizational needs or technical possibilities. Rather than choose one approach, we decided to extend their contributions by combining close-ended questions, minimal open-endedness in design, component recommendations, and iterative evaluation into one. We also add several system layers and components that were missing, to create a complete blueprint (based on design patterns, Gamma et al., 1994).

## 4    A Framework for DLT Utilization

Our analytical framework (cf. Figure 3) centers on three consecutive stages, formed as an iterative process for decision-making with clear inputs and outputs that are integrated across stages. Input represents the information that is documented in the answers, which leads to documented requirements as outputs. The 86 self-contained questions have close-ended outcomes and answer the three fundamental dilemmas behind every DLT-related prototyping effort: should I use DLT in my use case (**evaluation**); what general type of DLT access-based type is the best fit (**elicitation**); and how could the software architecture look like (**design**). Intended users are managers and software architects.

Each stage features three substages that hold the questions. The stages are intended to be as flexible as other workflows, but to include the strictness of decision trees and cost-benefit analyses of questionnaires at early points (cf. Bueno Momčilović et al., 2022). This approach is inspired by work in Balta et al. (2019), which also specified off-the-shelf tools to help answer the questions. For example, the Stakeholder Dependency Matrix is a tool for making stakeholder relationships explicit by diagramming the roles and interdependencies (Balta et al., 2015), which can help to crystalize the needs of users as design requirements. Such tools also provide the benefit of integrating multiple decisions, which we enable by having the output of one stage serve as partial forward and backward input for other ones. For instance, determining how business needs and technology properties align in the second stage could result in a decision against DLT, or bring an existing non-DLT legacy architecture into a blueprint of the intended DLT-based information system.

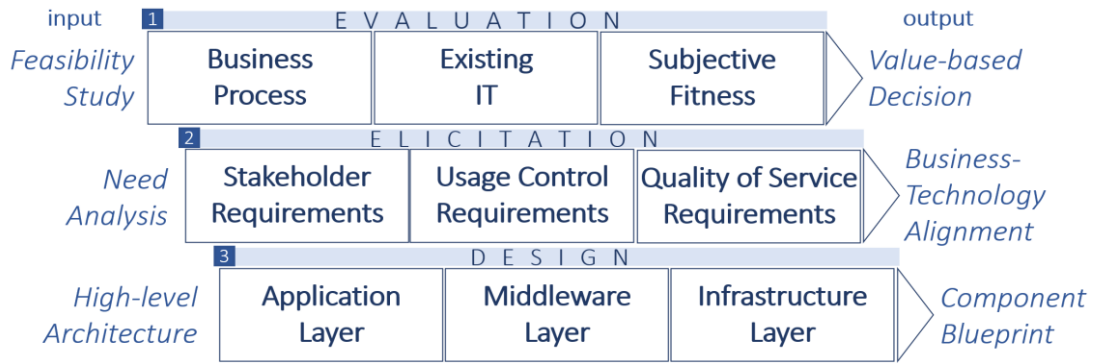

*Figure 3.    Towards traceable decisions for DLT utilization based on our framework*

In the first evaluation stage, users would evaluate the suitability of their use case by collecting their input through what we call a **feasibility study**. The goal is for managers to match the properties of the use case with the preconditions of using a DLT solution in the first two substages, and if there are no incompatible aspects (i.e., a need for deleting transactions versus using an immutable ledger), to understand potential risks in uncertain but workable mismatch in the third substage. **Business process** questions arise from the DLT properties that would unavoidably affect organizational processes, involving explicit measures for fixing misaligned stakeholder interests (cf. e.g. Mulligan et al., 2018) and finding ways towards 'trustless' relationships (cf. e.g. Maden & Alptekin, 2020).
**Existing IT** questions arise from the technical limitations of the DLT, including immutability of entries in the ledger (cf. e.g. Lo et al., 2017) or hardware-based requirements that may lead to upgrades (cf. e.g. Platt et al., 2021). Users are made aware of the constraints of e.g., limited throughput of most solutions (cf. e.g., Hribernik et al., 2020) and network scalability (cf. e.g. Koens & Poll, 2018). Finally, **subjective fitness** helps users evaluate whether DLT is still appropriate compared to alternatives. Users would need to decide whether certain benefits (e.g., avoiding censorship; Lapointe & Fishbane, 2019) are worth the risk and costs of using a potentially less fitting solution. The output is a **value-based decision**, which means that users have demonstrated or disproved the value of using DLT in their use case and decided on the path forward.
The second elicitation stage helps users to elicit detailed requirements from needs, which will later be translated into traceable design decisions. Managers and engineers (or architects) preferably elicit these together by specifying which questions are most important in the **need analysis**, guided by the questions. **Stakeholder requirements** questions center on the participants in the process (cf. e.g. Wüst & Gervais, 2017), their interactions (cf. e.g. Gourisetti, Mylrea & Patangia, 2019), and their incentives (cf. e.g. Degenhart et al., 2019; Habes et al., 2019). Users analyze who should determine consensus (cf. e.g. Maull et al., 2017), write transactions (cf. e.g. Birch, Brown & Parulava, 2016) and audit data (cf. e.g. Hunhevicz & Hall, 2020), among other examples.

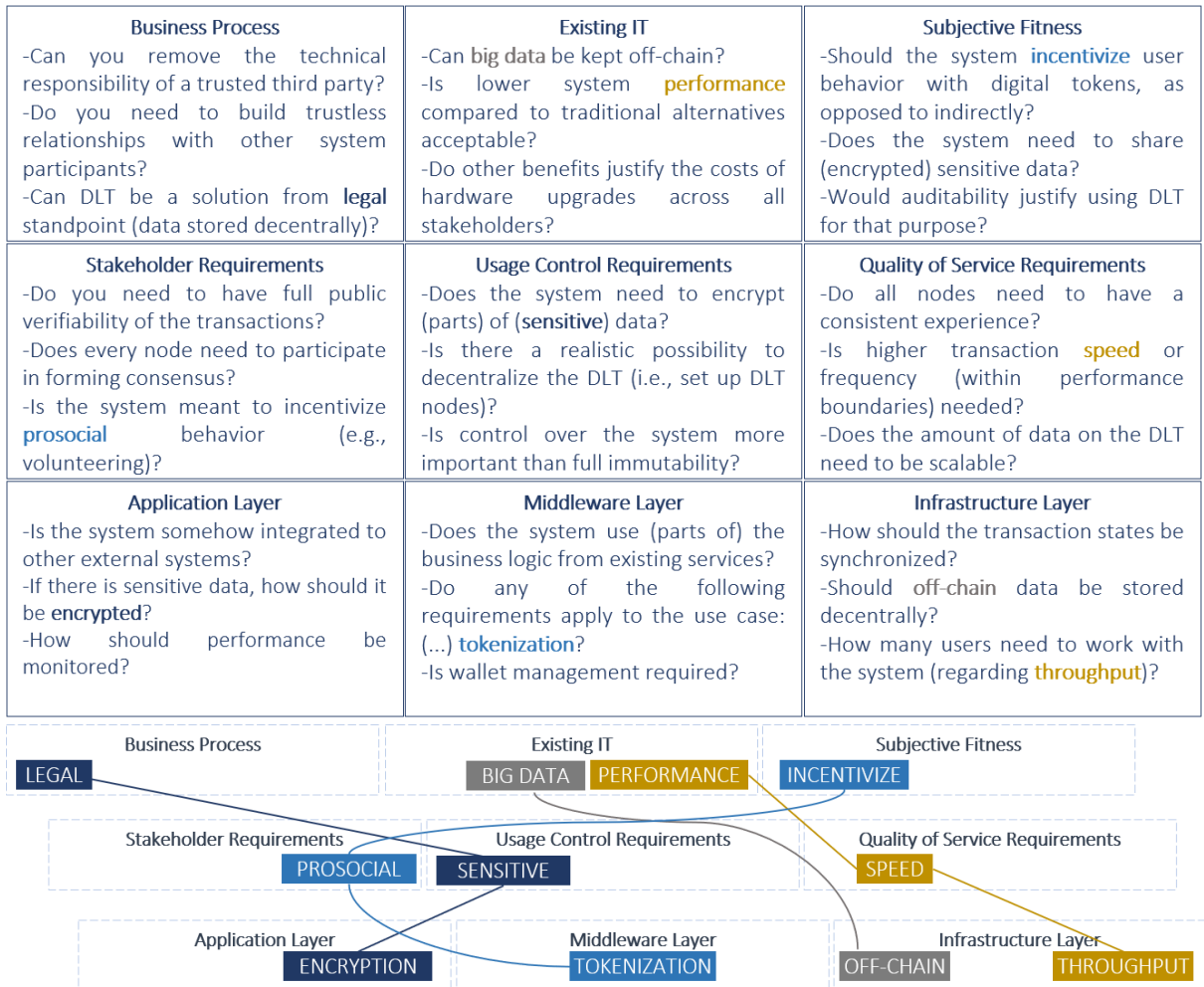

*Figure 4.    Example questions in each substage and their connections*

**Usage control requirements** address privacy or security reasons for compartmentalizing the flow of data and managing permissions (cf. e.g. Belotti et al., 2019), where e.g. private DLT solutions offer increased control at the expense of full public verifiability of transactions (cf. e.g. Mendhurwar & Mishra, 2021). **Quality of service requirements** provides initial points for on-chain scalability and performance (cf. e.g. CompTIA, 2019), to be later specified as on- and off-chain design components. Once the stage is completed and requirements are traced back to original needs, the output is a confirmation of **business-technology alignment**.

Finally, in the design stage, engineers translate the **high-level architecture** into a usable design blueprint. The application layer covers topics around data representation, while the middleware layer deals with data dissemination, and the infrastructure layer with data verification (Luo & Yan, 2021, Zhao et al., 2021). The **application layer** covers typical software application topics, such as the interaction with the system and users, devices, or web services, as well as the integration of relevant components with DLT and each other. It also includes classical system administration, such as monitoring or logging. **Middleware layer** covers DLT-specific functionality for users, including the codification of programmable business logic into smart contracts.

The **infrastructure layer** describes the backbone of the DLT-based system architecture through three sub-layers: network, processing, and storage. Connecting to the former stages, the network layer details the deployment of components, the design of consensus, scalability concerns, and security aspects. The storage layer deals with data management, specifically which data should be stored on-chain, which benefits from off-chain storage, and how the two can be linked. The overall output of architecture design is a suggested software architecture on component-level (i.e., **component blueprint**), that can be described from three distinct views based on user needs: a functional-engineering overview of the required Input, and covered aspects of the evaluated domain; process view describing the flow of information and interplay between single software components; and output view describing how to work with the output and what different users can benefit from.

Our online tool (https://dlt.fortiss-demo.org) is a web-based questionnaire representing the entirety of our framework. The inputs represent the implicit knowledge brought by users and recorded on the website – for which the tools for elicitation will be determined in future versions – and the outputs are generated on a static page with evaluation results, appropriate DLT types, and the component canvas. The goal of the tool is to allow users to revisit questions iteratively, generate different outcomes based on different use cases or conditions, save progress, and access a wiki to understand the terms and the reasoning behind the suggestions.

## 5 Exemplary Application

To demonstrate the application of our framework, we analyze a specific problem that was presented in the 2021 DLT4PI lab by a pharmaceutical consortium (cf. Table 2; Devarajan et al., 2021). Decentralizing and increasing the scope of clinical trials has been an ongoing challenge. The privacy of patients is the main requirement, accompanied by prohibitive legal consequences (e.g., GDPR) if their data is mismanaged. However, access to information on the effects of medication and the ability to maintain the quality of medication during the testing period is important for doctors and companies to avoid unknowingly creating poor or dangerous medication down the line. The goal is to prototype DLT to improve the coordination of the direct delivery of medication to patients, and track the trial while anonymizing patient data. Four stakeholder groups in these clinical trials include the pharmacological companies which develop the medicine for testing, the patients who are the trial subjects, the doctors who help conduct the trials, and the delivery organizations which distribute the medicine.

This application represents a simulated use of our analytical framework based on an existing challenge, its corresponding prototype, and documentation connecting the two. The input is collected using repositories and texts, but the output is generated manually and connected with the prototype (where such connections exist). This reflects the approach that a new user would take to understand the problem and find appropriate solutions with the blueprint and decisions they have reached using our tool.

| **Evaluation**: | **Business process** | **Existing IT** | **Subjective fitness** |
|---|---|---|---|
| decentralizing medically important clinical trials and helping ensure legal compliance. | trials have many contributors with misaligned interests that no third party can (legally or feasibly) manage alone. | shared immutable record from multiple writers are medically/ legally crucial, but not big data and high performance. | high fit; clinical trials benefit from 60% of properties, incl. sensitive data protection, decentralization, and auditing. |
| **Elicitation** | **Stakeholder req.** | **Usage control req.** | **QoS req**. |
| private network aligns with specified requirements of pharmaceutical companies and clinical trial stakeholders. | actors are many, not fully trusted, preselected, not incentivized on-chain, motivated by legal/ medical compliance and efficiency. | data needs to be encrypted, access pre-approved and compartmentalized, and process partially centrally overseen. | data and number of participants scale (feasibly) over time, so security and version compatibility are needed; high performance optional. |
| **Design** | **Application layer** | **Extension layer** | **Protocol layer** |
| Corda used as backbone DLT technology (private, permissioned) together with Java spring boot | Corda nodes called via RPC, REST-API, written in Java 8 (spring boot on tomcat web server), | Corda flows and contracts implement business logic in CorDapp. | Corda states for data storage and access, notary as network, timestamp and validate |

| middleware and React frontend. | frontend with React and Material UI. | | service, data sharing on need-to-know basis. |
|---|---|---|---|

*Table 2. Example application of the developed framework on the clinical trial case*

Our first stage involves evaluating whether DLT is a suitable solution for the case with 31 questions, which managers can answer using interviews and process mapping, with only small input from engineers. We rely on documentation from the lab, and find that the following criteria are fulfilled (among others): a need for 'trustless' verification due to largely misaligned incentives and interests of stakeholders; a willingness to remove central intermediaries, to improve efficiency and ensure GDPR-related legal compliance is met; involvement of multiple participants; and contract-based activities acting as transactions to be verified. Regarding existing IT, all technical constraints are acceptable. For example, immutable storage of anonymized medical data is needed so that anomalies (e.g., late deliveries, forgetful patients, or adverse reactions to the medication) can be traced. Logs of patient self-reports and deliveries need to be accessed by doctors and pharmaceutical companies, but only them, so even access to reading the transactions needs to be compartmentalized. Performance above DLT limitations and storage of big data on the ledger are not necessities, but a smartphone- and PC-compatible solution is, based on the hardware which doctors and patients have. Finally, by comparing the documentation with the subjective fitness questions, we understand that the needs to decentralize delivery, handle sensitive data and security of many patients, and audit the quality and compliance of each trial, are highly important qualities for the consortium, which together provide justifications for choosing DLT over other alternatives.

In the second stage, managers and engineers align the business needs with technical capabilities. The further 23 questions mimic a classical process of requirements engineering, but make the limitations and options with DLT-related properties explicit. Once again relying on documentation, the needs of our challenge translate into the following requirements: encryption of sensitive data; no coins, tokens, or transaction fees for users; no involvement of all stakeholders for consensus; decentralization of delivery transactions, but centralized read access for auditing; and long-term scalability through off-chain solutions. The primary incentives of the consortium are to simplify legal and medical compliance and increase process efficiency, to which a private, permissioned DLT solution corresponds. Nonetheless, the access control specifications are determined during design.

The final stage of 32 design questions is meant for engineers, with occasional input from management. We feed known requirements into technical specifications for components, and the resulting blueprint is a canvas of components. Four stakeholders need to interact with the system via a web app that uses an orchestration layer for preprocessing (e.g., anonymization) and communication with the DLT. Encryption of sensitive data and scalability need to be implemented with an off-chain storage solution. This fits with the idea that, for legal reasons, a private ledger solution can best separate who determines consensus and reads transactions by implementing strong identity and access management. If users realize that much larger scalability and performance might become an issue down the line, it would be useful to rethink using DLT or plan out mitigation steps. Lastly, to audit compliance more efficiently, handling files (e.g., images verifying that patients follow testing procedures) or integrating IoT devices (e.g., measurements that check medicine storage during delivery) is recommended.

# 6 Discussion

The exemplary application of our analytical framework shows how users can pinpoint specific paths forward in the highly contentious area of possibilities. It is meant to streamline the analysis of possible DLT use cases, and create an architecture blueprint for those where DLT is a promising solution. The framework is general enough to be adaptable to other problem domains without a one-size-fits-all approach (e.g., central bank digital currencies or energy prosumership; cf. fortiss, 2023). Hence, it can be directly beneficial to academia, the industry, and other sectors, to create benefits from changed processes, select technological components, or specify concrete requirements to match industrial demand, among other uses.

Our retrospective analysis during the development of the tool led to three challenges that future applications of the framework can help in solving: 1) verifying reasons for disqualification, 2) verifying recommended components, and 3) developing industry-specific branches. Regarding disqualification, some organizations behind the prototypes did not see certain properties to be incompatible to the extent that some authors of existing frameworks did. For example, many close-ended frameworks immediately disqualify the case if a trusted third party remains in the business process, to prevent users from wasting time or resources on primarily 'trustless' solutions (i.e., as a form of informed gatekeeping). However, organizations in the prototyping lab considered trustless solutions to complement third parties, as they can automate some (but not all) intermediary and more technical tasks that programs are suited to, and still provide room for third-party support in legal or administrative areas. Due to such contrasts between the two sources, we considered some questions to be softer limitations or risks that users need to be aware of. An agile and traceable process, with reduced gatekeeping in contested issues, could lead users to rethink their approach at a later point, or negotiate with their stakeholders towards better solutions.

Regarding recommended components, it became evident during the literature review that a holistic DLT decision framework recommending specific software architectures is yet to be developed. Design recommendations centered on evaluations of performance, implementations of single DLT use cases, or suggested reference architectures. While we address this gap, our recommendations of software architecture mainly apply to proofs-of-concept or minimum viable products. A potential extension might involve integrating extant research from Udokwu et al. (2021) and Xu et al. (2017), whose work primarily centers on more open decision support questions that may be complementary to analytical frameworks.

Regarding branches of the framework for specific industries, our approach is currently industry-agnostic, meaning that specific industry requirements or scenarios are neglected in favor of generalizability. This step of instantiating the framework in specific contexts will be pursued in the future and will likely impact the flexibility and level of detail of our method across other problem domains. Moreover, the technologies and frameworks we recommend in the architecture design are on a component level and mostly multiple suggestions are provided. The reason is that in most cases, the specific implementation and detailed architecture on a lower level is often up to the preferences of an organization or individual. Consequently, specific advantages and disadvantages of a single technology or framework are beyond the scope of current research.

# 7    Conclusion

We develop a holistic framework that covers whether DLT should be used and what a suitable software architecture could look like. By applying it, managers and software architects can develop suitable architectures more quickly and completely than it was earlier possible, and re-evaluate existing ones. Ultimately, our framework drives users to utilize DLT in a structured manner and strives to combine business and technology perspectives holistically.

Still, two general limitations remain: (i) tools and details for iterating through the process are not specified, and (ii) the framework was not directly evaluated by experts or practitioners. First, the framework guides users towards important questions, but does not provide a corresponding toolkit to answer them. Our contribution is currently general, and applicability alongside various existing methods (e.g., requirements engineering) or applications (e.g., Hyperledger Fabric) is also untested, when mapping stakeholder incentives or researching off-chain components, for example. Second, questions from our meta-study were not validated by experts, thus making the details tentative in terms of their real value. Although prototypes provide some information from a sample of organizations and students, our future aim is to use expert input to ensure that each question is necessary, and that contradictions between theory and practice are resolved.

Future research will focus on addressing the stated limitations and other aspects. Detailed recommendations, for example, could involve more specific components, technologies, and methods structured along usable DLT blueprints, by making suggestions that fit each industry, but also avoiding advice that may become obsolete in the future. Similarly, more agile approaches involving groups of stakeholders may be relevant in managing change alongside stakeholders with different decision-making power. Nonetheless, we believe that our preliminary results provide clear benefits for practice and will steer a vivid discussion in academia towards a more evidence-based use of DLT across industries and beyond hype cycles.

## Acknowledgments

This work was partially funded by the following research projects: BayernCloud (20-13-3410.1-01A-2017) funded by the Bavarian Ministry of Economics as well as BEST (03EI4017C), DiProLeA (02J19B122) and K4R (01MK20001F) funded by the German Ministry of Economics. Moreover, the research would not have be able to materialize without the practical course at TU Munich organized by team at the chair of Professor Helmut Krcmar. Finally, we received detailed reviews from three reviewers. We are grateful for all the support.